\documentclass[5p]{elsarticle}
\usepackage{amsmath}
\usepackage{amssymb}
\newcommand{\dd}{\, {\rm d}}

\usepackage{cancel}
\usepackage{bigstrut}
\newcommand{\lsim}{\;\mbox{\raisebox{-0.5ex}{$\stackrel{<}{\scriptstyle{\sim}}$}
}\;}
\newcommand{\mpl}{M_{\mathrm{pl}}}
\newcommand{\gp}{{g^\prime}}
\usepackage{subfigure}
\newcommand{\gsim}{\;\mbox{\raisebox{-0.5ex}{$\stackrel{>}{\scriptstyle{\sim}}$}
}\;}

\newcommand{\pmi}{\phi_{\rm min}}
\newcommand{\rc}{\rho_{\rm c}}
\newcommand{\bv}{\beta_\varphi}

\begin{document}
\begin{frontmatter}
\title{SUPER-Screening}
\author[pb]{Philippe Brax}
\ead{Philippe.Brax@cea.fr}
\author[damtp]{Anne-Christine Davis}
\ead{A.C.Davis@damtp.cam.ac.uk}
\author[damtp]{Jeremy Sakstein}
\ead{J.A.Sakstein@damtp.cam.ac.uk}
\address[pb]{Institut de Physique Theorique, CEA, IPhT, CNRS, URA 2306, F-91191Gif/Yvette Cedex, France}
\address[damtp]{DAMTP, Centre for Mathematical Sciences, University of Cambridge, Wilberforce Road, Cambridge CB3 0WA, UK}

\begin{abstract}
We present a framework for embedding scalar-tensor models of screened modified gravity such as chameleons, symmetrons and environmental dilatons into global supersymmetry. This achieved by secluding the dark sector from both the observable and supersymmetry breaking sectors. We examine the resulting supersymmetric features in a model-independent manner and find that, when the theory follows from an underlying supergravity, the mediation of supersymmetry breaking to the dark sector induces a soft mass for the scalar of order the gravitino mass. This is enough to forbid the construction of supersymmetric symmetrons and ensures that when other screening mechanisms operate, no object in the universe is unscreened thereby precluding any observable signatures. In view of a possible origin of modified gravity within fundamental physics, we find that no-scale models are the only ones that can circumvent these features. We also present a novel mechanism where the coupling of the scalar to two other scalars charged under $\mathrm{U}(1)$ can dynamically generate a small cosmological constant at late times in the form of a Fayet-Iliopoulos term.
\end{abstract}
\end{frontmatter}

\section{Introduction}

Dark energy is one of the simplest and most plausible explanations of the recent discovery of the acceleration of the expansion of the Universe \cite{Perlmutter:1998np,Riess:1998cb}. Unfortunately, this very simple theory is fraught with difficulties. The cosmological constant problem is by far the most famous, with no current understanding of the tiny value of the dark energy scale compared to other energy scales inherent in particle physics, which has prompted the search for alternative sources of vacuum energy in the form of slowly-rolling quintessence fields \cite{Copeland:2006wr}. Fine-tuning and coincidence problems aside, the mass of the field must be small ($\mathcal{O}(H_0)$) on cosmological scales and so any coupling to matter, which one would naturally expect, results in long-ranged \textit{fifth} forces, which are not compatible with solar-system tests of gravity.

Any such coupling of a scalar field to matter is equivalent to a low-energy modification of general relativity (GR) \cite{Weinberg:1965rz} and, in particular, the fifth-force problem has led to the development of screening mechanisms (see \cite{Jain:2010ka} for a review) where fifth-forces are screened locally, on galactic or solar system scales, but are active over large, cosmological scales. All of our current experimental tests of gravity have been performed in our local neighbourhood and so there is nothing precluding this possibility. Scalar-tensor theories, where a new scalar degree of freedom couples conformally to the metric, are one such class of models and can screen fifth-forces in high density environments either through the chameleon mechanism \cite{Khoury:2003aq,Khoury:2003rn,Brax:2004qh}, where the local mass of the field is large enough that the range of the fifth-force is sub-mm, or the symmetron \cite{Hinterbichler:2010es} and environmental screening \cite{Brax:2010gi} models, where the strength of the fifth-force is rendered negligible. Whilst the concept of a classical screening mechanism is robust there are two drawbacks: a cosmological constant is still required to account for the acceleration of the universe \cite{Wang:2012kj} and it is unclear whether the mechanism is stable to quantum corrections (see \cite{Upadhye:2012vh} for a discussion on the one-loop effects and \cite{Gannouji:2012iy} for a discussion of the quantum properties of such theories).

Scalar vacuum expectation values (VEVs) are unstable to quantum corrections, which makes finding a natural value for the cosmological constant difficult in quintessence-like scenarios. One solution to this is to impose mildly broken supersymmetry, which produces only small quantum corrections when the system is close to a supersymmetric vacuum. Particle physics in such theories operates at the TeV scale and a lot of attempts have been put forward to solve the problem of corrections of this magnitude by decoupling the observable and dark sectors so that they interact only weakly between themselves and with  a hidden supersymmetry breaking sector. One may then hope that both of the above issues can be ameliorated by constructing a supersymmetric extension of screened modified gravity theories with decoupled sectors.

Here we shall do exactly this. Scalar-tensor theories are low-energy IR modifications of GR, which are (at best) valid at energy scales below big bang nucleosynthesis (BBN) and so rather than focus on supergravity \cite{Brax:2006kg,Brax:2006dc,Brax:2006np} or string theory \cite{Hinterbichler:2010wu,Conlon:2010jq}, both of which have had issues with no-go theorems or spatial decompactifications we will build on the approach of \cite{Brax:2011qs} and construct a general $\mathcal{N}=1$ globally supersymmetric framework for describing these theories. A chiral scalar field couples to dark matter fermions in a sector completely secluded from the observable one resulting in an enhancement of the gravitational force that is screened on small scales. Specific models may be realised within this framework via different choices of the K\"{a}hler potential, superpotential and  coupling function, although, with the exception of illustrative examples, model-dependent results will be presented in \cite{future_susy}.

We investigate the new features introduced by supersymmetry. Supersymmetry is always broken at finite densities, however the scale is density dependent and, owing to the seclusion of the dark sector, far lower than that in the observable sector. When the theory derives from an underlying supergravity that is broken at some high energy scale, there is correction to the scalar field's mass of order the gravitino mass $m_{3/2}$. This constraint immediately precludes the possibility of embedding symmetron theories into supersymmetry unless one is willing to tolerate excessive fine-tuning. Chameleon and dilaton models survive, however the constraint is so strong that it ensures that no object in the universe is unscreened and it is therefore not possible to detect supersymmetric screened modified gravity observationally using fifth-force effects\footnote{There is still the possibility of using a coupling to photons, however we will not investigate this here.}. Only no-scale K\"{a}hler potentials with a certain isometry group for the scalar manifold can evade these constraints. These models are therefore strong candidates for finding screened modified gravity in more UV complete theories such as string theory and supergravity and any future work in searching for these theories should concentrate on this class of models. On the other hand it is well-known that it is difficult to construct chameleon theories using this type of K\"{a}hler potential and so instead we illustrate these new results by constructing a class of models which possess the chameleon mechanism and a supersymmetric vacuum, postponing a full analysis of their dynamics to later \cite{future_susy}.

By coupling the field to two $\mathrm{U}(1)$ charged scalars, our model can generate a natural cosmological constant in the form of a Fayet-Illiopoulos (FI) term. The charged scalar potential is $\mathrm{U}(1)$ symmetric and at early times the symmetry is broken. The coupling to the gravitational scalar gives a field-dependent contribution to the charged scalars masses and as the cosmological field evolves along its density dependent minimum this mass increases until some critical density where it vanishes and the symmetry is restored, leaving only the FI term, which acts as a cosmological constant. FI terms are not renormalised when supersymmetry is unbroken and otherwise runs only logarithmically. Therefore this cosmological constant does not receive quantum corrections. Our low-energy description is not powerful enough to fix the value of the FI constant and we must tune it to $10^{-3}$ eV in order to account for the energy density in dark energy, however, if one could find a natural reason for such a small value in a more UV complete theory, for example a suitable ratio of different mass scales, this value would not be corrected and would persist all the way down to present-day energy densities. Throughout this work we will assume that the cosmological constant problem in the observable sector is resolved and focus only on the dark sector.

In the next section we review scalar-tensor modified gravity and introduce the general framework for embedding it into global supersymmetry. In section (\ref{sec:features}) we describe the new features which this framework adds to these models which we illustrate in section (\ref{sec:modelb}) by constructing a class of supersymmetric chameleons. In section (\ref{sec:cc}) we show how we can incorporate a cosmological constant using a hybrid mechanism before concluding in section (\ref{sec:concs}).

\section{Supersymmetric Screened Modified Gravity}\label{sec:MG}

\subsection{Scalar-Tensor Screened Modified Gravity}

The action
\begin{align}\label{eq:stmod}
 S&=\int\dd^4x\sqrt{-g}\left[\mpl^2\frac{R}{2}-\frac{1}{2}k^2(\phi)\nabla_\mu\phi\nabla^\mu\phi-V(\phi)\right.\nonumber\\ &+\left.\vphantom{\mpl^2\frac{R}{2}}\mathcal{L}_{\rm m}(\Psi_i;g_{\mu\nu})+\mathcal{L}_{\rm c}(\chi_i;A^2(\phi)g_{\mu\nu})\right]
\end{align}
describes a scalar coupled minimally to matter $\Psi_i$ but non-minimally to cold dark matter $\chi_i$ via the Weyl rescaled metric $\tilde{g}_{\mu\nu}=A^2(\phi)g_{\mu\nu}$; $A(\phi)$ is known as the \textit{coupling function}. This function satisfies $A(\phi)\approx1$ so that the perturbations with respect to each metric, $g$ or $A^2g$, are small, which motivates the heuristic form $A(\phi)=1+\mathcal{O}(\phi/M)+\ldots$. This coupling defines a two-component fluid, the scalar and dark matter, which can exchange energy and so the energy density defined by $\rho=-T$, $T$ being the trace of the energy-momentum tensor found using $g_{\mu\nu}$, is not conserved. Instead, it is the energy-momentum tensor defined using $\tilde{g}_{\mu\nu}$ which is covariantly conserved, however it can be shown that the rescaled energy density $\rho_c$ where
\begin{equation}
\rho=A(\phi)\rc;\quad \rho=-T_{}=-g_{\mu\nu}T_{\rm m}^{\mu\nu};\quad T^{\mu\nu}_{m}=-\frac{2}{\sqrt{-g}}\frac{\delta S_{\rm c}}{\delta g_{\mu\nu}},
\end{equation}
satisfies the continuity equation and so is conserved non-relativistically. We shall henceforth refer to $\rc$ as the conserved dark matter density. Dark matter particles move along geodesics of $\tilde{g}$, the so called \textit{Jordan Frame} metric and not $g$, the \textit{Einstein Frame} metric. Observers in the Einstein frame therefore infer an additional or \textit{fifth} force
\begin{equation}\label{eq:fifthforce}
 {\bf F_{\phi}}=\frac{\beta_\varphi(\varphi)}{\mpl}{\bf \nabla}\varphi; \quad \beta_\varphi(\varphi)\equiv\mpl\frac{\dd\ln A(\varphi)}{\dd \varphi},
\end{equation}
where $\varphi$ is the canonically normalised field; $\dd\varphi=k(\phi)\dd\phi$. The equation of motion for the field is
\begin{equation}\label{eq:eom}
 \Box\varphi=\frac{\dd V_{\rm F}(\varphi)}{\dd \varphi}+\rho_c\frac{\dd A(\varphi)}{\dd \varphi},
\end{equation}
which is the usual Klein-Gordon equation with an effective potential
\begin{equation}\label{eq:effpotst}
 V_{\rm eff}(\varphi)=V_{\rm F}(\varphi)+\rc (A(\varphi)-1).
\end{equation}
The final term arises due to the coupling to dark matter.

Models such as these are known to possess \textit{screening mechanisms} \cite{Khoury:2003aq,Khoury:2003rn,Brax:2010kv,Hinterbichler:2010es,Brax:2010gi,Brax:2008hh,Brax:2012gr} (see \cite{Khoury:2010xi} for a review) where the fifth-force is rendered negligible in dense environments. This feature requires two properties: the effective potential must possess a minimum and either the mass at this minimum must be very large such that the force is Yukawa suppressed (this is known as the \textit{chameleon} mechanism \cite{Khoury:2003rn}) or the coupling $\beta_{\varphi}(\varphi)$ must become small enough that the force is negligible (this is the mechanism employed by the symmetron \cite{Hinterbichler:2010es} and the environmentally dependent Damour-Polyakov effect \cite{Brax:2010gi}). In this paper we shall work mainly with chameleon theories, referring briefly to the symmetron whilst discussing supergravity corrections, however many of our results apply equally to all three models. We will indicate where this is the case.

\subsection{Supersymmetric Scalar-Tensor Theories}\label{sec:susyst}

The theory described above can be realised within a supersymmetric framework by coupling a chiral superfield $\Phi=\phi+\ldots$, whose lowest component plays the role of the scalar, to two other fields $\Phi_\pm=\phi_\pm+\sqrt{2}\theta\psi_\pm+\ldots$, whose fermions act as dark matter. The K\"{a}hler potential is
\begin{equation}
 K(\Phi,\Phi^\dagger,\Phi_\pm,\Phi_\pm^\dagger)= \Phi_+\Phi_+^\dagger+\Phi_-\Phi_-^\dagger+\hat{K}(\Phi,\Phi^\dagger),
\end{equation}
where $\hat{K}(\Phi,\Phi^\dagger)$ is left unspecified for now; its specific form is crucial for determining which screening mechanism is utilised or indeed if one is even present. When $\hat{K}(\Phi\Phi^\dagger)\ne \Phi^\dagger\Phi$ the field is not canonically normalised, indeed
\begin{equation}
 \mathcal{L}_{\rm kin}\supset K_{\Phi\Phi^\dagger}\nabla_\mu\Phi\nabla^\mu\Phi^\dagger
\end{equation}
so that the mass of the field is
\begin{equation}\label{eq:Phimass}
 m^2_\Phi=\frac{1}{K_{\Phi\Phi^\dagger}}\frac{\partial^2V(\Phi)}{\partial\Phi\partial\Phi^\dagger}.
\end{equation} The superpotential is
\begin{equation}\label{eq:suppotgen}
 W(\Phi,\Phi_\pm)=\hat{W}(\Phi)+mA(\Phi)\Phi_+\Phi_-.
\end{equation}
Again, we leave $\hat{W}$ unspecified since a specific choice of its form leads to different models. With this arrangement, $\langle\phi_+\rangle=\langle\phi_-\rangle=0$ and so the potential is
\begin{equation}\label{eq:VFpot}
 V(\Phi)=V_{\rm F}(\Phi)=\hat{K}^{\Phi\Phi^\dagger}\left\vert\frac{\dd \hat{W}}{\dd \Phi}\right\vert^2
\end{equation}
whilst there is a $\Phi$-dependent contribution to the dark matter fermion mass
\begin{equation}
 \mathcal{L}_{\rm f}=\frac{\partial ^2 W}{\partial\Phi_+\partial\Phi_-}\psi_+\psi_-=mA(\Phi)\psi_+\psi_-.
\end{equation}
When these fermions condense to finite density such that $\langle\psi_+\psi_-\rangle=\rc/m$ this term provides an additional contribution to the potential resulting in an effective potential
\begin{equation}\label{eq:veffPhi}
 V_{\rm eff}(\Phi)=V_{\rm F}(\Phi)+\rc(A(\Phi)-1).
\end{equation}
This  model is therefore equivalent to the model presented in equation (\ref{eq:stmod}). In practice, it will be necessary to decompose $\phi$ as $\phi=|\phi|e^{i\theta}$ and stabilise the angular field at the minimum, however several general results can be derived before specialising to specific models and so we shall continue to work with $\Phi$ for the time being. When this decomposition is used we shall set $\phi\equiv |\phi|$ and use $\varphi$ to denote the field found by bringing the kinetic term for $\phi$ into canonical form.

\subsection{Supersymmetric Features}\label{sec:features}

In this section we will discuss some of the new features that accompany the embedding of these theories into a supersymmetric framework.

\subsubsection{Supergravity Corrections}\label{sec:sugracorr}

When working in the low-energy framework of global supersymmetry it is important to ensure that any corrections coming from supergravity breaking in a hidden sector are negligible. The most important correction for these models are those coming from $|D_\Phi W|^2$ of the form
\begin{equation}\label{eq:sugracorr}
 \Delta V_{\cancel{\rm SUGRA}}=\frac{K^{\Phi\Phi^\dagger}|K_\Phi|^2|W|^2}{\mpl^4}e^{\frac{K}{\mpl^2}}=m_{3/2}^2K^{\Phi\Phi^\dagger}|K_\Phi|^2,
\end{equation}
 where $m_{3/2}$ is the gravitino mass. This correction must be negligible compared to $V_{\rm F}$ and $\rc A(\Phi)$ if they alone are to be responsible for the screening mechanism \footnote{If this is not the case then one is really working within the framework of supergravity and can therefore not realise any screening mechanisms due to the no-go result of \cite{Brax:2006np}.}. This correction introduces an important new feature into these models: the mass of the field is always at least as large as the gravitino mass. To see this, one can take derivatives of (\ref{eq:sugracorr}) and focus on certain terms only to find
\begin{equation}\label{eq:m32massres}
 \frac{\partial^2V(\Phi)}{\partial\Phi\partial\Phi^\dagger}\supset m_{3/2}^2K_{\Phi\Phi^\dagger}.
\end{equation}
Recalling that the field may not be canonically normalised and applying (\ref{eq:Phimass}) one finds that there is a contribution to the field's mass of exactly $m_{3/2}$. This can be anywhere from $1$ eV as predicted by gauge mediated supersymmetry breaking scenarios to $\mathcal{O}(\textrm{TeV})$ corresponding to gravity mediated breaking \cite{Nilles:1983ge}. Consequently, the Compton wavelength of the field is $\lambda_{\rm c}\sim m_{3/2}^{-1}$ and so the range of the fifth-force in such models is always less than $10^{-6}$ m depending on the gravitino mass. It should be noted that this result is completely independent of the form of the matter coupling or the potential, it is not even sensitive to their origins or whether the field is coupled to dark matter or the standard model. When one has scalars coupled to matter and the theory has an underlying $\mathcal{N}=1$ supergravity at some high energy scale then the range of the fifth-force will always be less than $m_{3/2}^{-1}$. In supergravity breaking scenarios with a large gravitino mass this precludes the need for screening mechanisms  altogether.

\subsubsection{A Supersymmetric Symmetron}

Another immediate consequence of this is that canonical symmetrons \cite{Hinterbichler:2010es} cannot be accommodated within a supersymmetric framework. The supersymmetric symmetron is found by imposing a $\mathbb{Z}_2$ symmetry upon the effective potential. This is achieved by including only odd powers of $\Phi$ in $\hat{W}(\Phi)$ and only even powers in the coupling $A(\Phi)$. The K\"{a}hler potential is $\hat{K}(\Phi\Phi^\dagger)=\Phi^\dagger\Phi$ so that the fields are canonically normalised and, at lowest relevant order, the superpotential is
\begin{equation}\label{eq:suppotss}
 W(\Phi)=M^2\Phi+\frac{1}{3}g\Phi^3+m\left(1-h\frac{\Phi^2}{2m\Lambda_3}+f\frac{\Phi^4}{4m\Lambda_3^3}\right)\Phi_+\Phi_-,
\end{equation}
where the explicit introduction of the $-$ sign in the coupling will become clear momentarily and is completely consistent with supersymmetry. The F-term potential is then
\begin{equation}\label{eq:sspot1}
 V_{\rm F}(\phi,\theta)=M^4+g^2\phi^4+2gM^2\phi^2\cos(2\theta),
\end{equation}
which is minimised when $\cos(2\theta)=-1$ (one can check that if one includes a term of order $\Phi^5$ in (\ref{eq:suppotss}) then this is still approximately the case) so that the model is a symmetron:
\begin{align}\label{eq:supersymmv}
 V(\phi)&=M^4 -2gM^2\phi^2+g^2\phi^4=(g\phi^2-M^2)^2,\nonumber\\A(\phi)&=1+h\frac{\phi^2}{2m\Lambda_3}+f\frac{\phi^4}{4m\Lambda_3^3}.
\end{align}
 Note that this has a supersymmetric minimum ($V=0$) at $\phi_0=M/\sqrt{g}$; at finite density the field moves to smaller value thereby breaking this supersymmetry. Now the symmetron mechanism requires that the bare mass be negative, however there is a contribution from supergravity corrections of the order $+m_{3/2}^2$ and so either we must demand that there is a fine-tuned cancellation or we must take $M>m_{3/2}$ (note that the canonical symmetron model requires $M\le10^{-29} eV$ \cite{Hinterbichler:2010es,Hinterbichler:2011ca} whereas $m_{3/2}\ge 1$ eV). If this is not the case the symmetron mechanism is lost. Suppose then that $M>m_{3/2}$. We have
\begin{equation}
 \beta(\phi)=\mpl\frac{\dd \ln A(\phi)}{\dd \phi}\sim \frac{\mpl\phi_0}{m\Lambda_3}
\end{equation}
in the cosmological background and when the $\mathbb{Z}_2$ symmetry is broken (if this is not the case there is no fifth-force). So if the force is to be of comparable strength to gravity in free space we need $M\mpl\sim m\Lambda_3$. Now the symmetry is broken (or restored) at a density
\begin{equation}
 \rho_\star\sim M^2m\Lambda_3\sim M^3\mpl>m^3_{3/2}\mpl>10^{27}\textrm{ eV}^{4}=10^{39}\rho_0,
\end{equation}
where we have taken the best case scenario of an eV mass gravitino. This means that in the late-time universe only objects whose densities exceed $10^{27}$ eV$^4$ (and in many models vastly exceed) can restore the $\mathbb{Z}_2$ symmetry locally and screen the fifth-force. This immediately precludes screening in all dark matter haloes (with greatest density $10^6\rho_0$) and Earth based laboratories (with density $10^{29}\rho_0$). This problem is not ameliorated if we instead allow the force in free space to be stronger than gravity since this increases the lower bound on $\rho_\star$. Either the symmetron mechanism does not exist or it is unscreened in our solar system.

One may then also wonder whether the same is true of generalised symmetrons \cite{Brax:2012gr}. These employ a similar mechanism to canonical symmetrons, except the phase transition, which is still second order, is associated with a term which is higher order than quadratic. For example, the effective potential
\begin{equation}
 V_{\rm eff}(\phi)= -g\phi^4+\frac{\phi^6}{\mathcal{M}^2}+\cdots+\rho\frac{\phi^4}{\mu^4},
\end{equation}
where all higher order terms in the potential have positive coefficients, has a broken $\mathbb{Z}_2$ symmetry which is restored in dense enough environments. In this case, a correction to the mass of the order $m_{3/2}$ does not affect the choices for the parameters in the theory, however it does add a term proportional to $m_{3/2}^2\phi^2$ to the effective potential and hence changes the transition from second to first order. In this case, the original mechanism is lost and $\beta(\phi)$ does not approach zero smoothly in increasingly dense environments. The properties of first order transitions are notably different from those which are second order and it is unclear whether any screening mechanism can persist once this term is included. In the reconstruction of the potential from the coupling function and effective mass investigated by \cite{Brax:2012gr} such quadratic terms were not allowed.

\subsubsection{No-Scale-Type Models}

Given that the mere presence of an underlying supergravity imposes such stringent restrictions on the mass of the field one might naturally wonder how general these restrictions really are and whether they can be circumvented. There are indeed a class of models where the supergravity correction, i.e. the mass (\ref{eq:m32massres}) is not present. Clearly if $K^{\Phi\Phi^\dagger}|K_\Phi|^2$ is constant then (\ref{eq:m32massres}) is spurious since the second derivative of the corrections are zero and there are no corrections to the field's mass. These are the no-scale type models, a particularly common example of which is the logarithmic K\"{a}hler potential that arises in type IIB superstring theory $K=-n\mpl^2\ln[(\Phi+\Phi^\dagger)/\mpl]$ ($n=1$ for the dilaton and $n=3$ for T-moduli, which corresponds to the pure no-scale case). In more complicated scenarios one typically has many chiral scalars, which parametrise a no-scale type manifold given by $K^{\Phi_i\Phi_j^\dagger}K_{\Phi_i}K_{\Phi_j^\dagger}=c$ with $c=3$ in the pure no-scale case.

At tree-level, these models evade the corrections, however one may wonder if they are re-introduced by loop corrections. The one-loop effective potential is
\begin{equation}\label{eq:1loopV}
 \Delta V_{\rm 1-loop}=\frac{1}{64\pi^2}\textrm{STr}\left[M^4\ln\frac{M^2}{\mu^2},\right]
\end{equation}
where $M$ is the mass matrix and $\mu$ is the renormalisation group scale. At tree level we have, for the scalar, $M^2\sim |\hat{W}_{\Phi\Phi}|^2$ and so if $\hat{W}\sim \mathcal{O}(\mathcal{M}^3)$ there we expect $\mathcal{M}\ll\mpl$ since $\hat{W}$ is associated with low-energy behaviour well below the supergravity breaking scale. In this case, the quantum corrections are set entirely by the tree-level parameters, which a priori are independent of the gravitino mass.

The equation (\ref{eq:1loopV}) encompasses only supersymmetric corrections and so we must also account for the supersymmetry breaking soft masses induced by this process. These have been studied extensively by \cite{Brignole:1993dj,Brignole:1997dp,Farquet:2012cs} (and references therein) who find that whenever the manifold is not pure no-scale i.e. $K^{\Phi_i\Phi_j^\dagger}K_{\Phi_i}K_{\Phi_j^\dagger}\ne3$ the soft masses are always of order $m_{3/2}$ and so one can conclude that these models do not  evade the supergravity breaking constraints. Furthermore, in the pure no-scale case the same analyses have shown that only no-scale models where the isometry group of the scalar manifold is
\begin{equation}\label{eq:manifod}
 \mathcal{M}=\frac{\mathrm{SU}(1,n)}{\mathrm{U}(1)\times\mathrm{SU}(n-1)},
\end{equation}
do not acquire soft masses. Any no-scale model whose isometry group differs from this must necessarily include gravitino-mass scalars in its low-energy effective theory.

One should note however that it is very difficult to find screening mechanisms with this class of K\"{a}hler potentials. For example, in the simplest case where $K=-3\mpl^2\ln(T+T^\dagger)/\mpl$, the canonically normalised field is $\Phi={\rm exp}(\sqrt{2/3}\phi/\mpl)$ and so we expect these to give rise to the chameleon mechanism. Now any term in the superpotential is of the form $W(\Phi)\propto \Phi^n$ and so at best one has an exponentially decreasing scalar potential and it is very difficult to obtain a thin shell solution for an Earth-like density profile in such a model \cite{Brax:2010gi}. One must then rely on non-perturbative effects to generate a viable potential. It has been shown in \cite{Hinterbichler:2010wu} that a non-perturbative superpotential $W\propto \textrm{exp}(aT)$ arising from gaugino condensation coupled to the KKLT mechanism \cite{Kachru:2003aw} can give rise to a chameleon. The potentials found using the standard string theory prediction tend to have the opposite effect and decompactify the extra dimensions \cite{Conlon:2010jq}.

We can then discern the conditions under which globally supersymmetric theories are not bound by constraints from supergravity breaking; they must be no-scale models with the isometry group (\ref{eq:manifod}). At the level of string theory, models such as these receive corrections to their K\"{a}hler potentials in string perturbation theory, which are then used in the tree-level supergravity formula to find the scalar potential. Hence, only string theory models which preserve this no-scale property to all orders in perturbation theory and under non-perturbative corrections can evade the supergravity correction to the mass. At the level of pure-field theory, any no-scale model with this isometry group will always lead to a low-energy model which is not bound by these constraints.

\subsubsection{Efficient Screening}

When the models do not evade the supergravity corrections the presence of a contribution of order $m_{3/2}$ to the field's mass is enough to ensure that the screening in these models is so efficient that no object in the universe is unscreened and it is impossible to measure the effects of any fifth-forces. This can be deduced as follows. Working with the canonically normalised field $\varphi$ and assuming that $\theta$ is stabilised at its minimum, the new effective potential is
\begin{equation}
 V_{\rm eff}(\varphi)=V_{\rm F}(\varphi)+\rho (A(\phi)-1)+\frac{1}{2}m_{3/2}^2\varphi^2.
\end{equation}
At the minimum, we have
\begin{equation}
 -\frac{\beta(\varphi_0)\rho_0}{\mpl\varphi_0}= m_{3/2}^2+\frac{1}{\varphi_0}\frac{\dd V_{\rm F}(\varphi)}{\dd \varphi}
\end{equation}
where we have used the fact that $A(\varphi_0)\approx1$. If this equation has no solutions then there is no minimum and the theory is not one of screened modified gravity. We are interested in situations where this is not the case and so we will assume that a minimum exists. Now the left hand side of this equation, barring any fine-tuned cancellation, must be as large as $m_{3/2}^2$, in which case
\begin{equation}\label{eq:betcond}
 \frac{\varphi_0}{\bv(\varphi_0)}\le\frac{\rho_0}{m_{3/2}^2\mpl}.
\end{equation}
It is well known that an object is self-screened when the parameter (subscript 0's refer to present day cosmological values)
\begin{equation}\label{eq:chi0def}
 \chi_0\equiv\frac{\varphi_0}{2\mpl\bv(\varphi_0)}
\end{equation}
 is less than the Newtonian potential $\Phi_{\rm N}=GM/R$ at the surface of the object\footnote{For a nice discussion of $\chi_0$, its relation to the Newtonian potential and the self screening of objects see \cite{Hui:2009kc}. A full derivation of this condition can be found in \cite{Hui:2009kc,PhysRevD.85.123006,Brax:2012gr}.}. When coupling to dark matter only we require that the dark matter halo of the Milky Way is self-screening\footnote{This condition may be relaxed if one instead demands that the milky way is screened by its local group partners. This raises the bound to $\chi_0<10^{-4}$, however, a recent, independent constraint using water maser distances precludes values $\gsim 10^{-6}$ thereby negating this argument. This case does not need to be considered here since we shall see that the self screening of the milky way is inherent in our model and does not require additional constraints} so that $\chi_0<10^{-6}$. Whereas this is the case with our simple model of section (\ref{sec:susyst}), the results of the previous analysis do not change if one were to couple to standard model (or beyond) particles, in which case compatibility with astrophysical tests such as \cite{Chang:2010xh,PhysRevD.85.123006,Jain:2012tn} require $\chi_0<5\times10^{-7}$.

Taking equation (\ref{eq:betcond}) and inserting it into (\ref{eq:chi0def}) we find
\begin{equation}\label{eq:chicond}
\chi_0\le \left(\frac{H_0}{m_{3/2}}\right)^2,
\end{equation}
where $\rho_0\sim 3\Omega_{\rm c}^0H_0^2\mpl^2$. In the best case scenario we have $m_{3/2}\sim\mathcal{O}(\textrm{eV})$ and so we have $\chi_0\le10^{-66}$. The most unscreened objects in the universe are dwarf galaxies with $\Phi_{\rm N}\sim 10^{-8}$ \cite{Hui:2009kc,Jain:2011ji} and so this condition ensures that no object in the universe is unscreened; hence, there are no observational signatures of fifth-forces.

This behaviour can alternatively be seen by considering the usual derivation of the unscreened field profile \cite{PhysRevD.85.123006,Brax:2012gr} in the unscreened region of a spherical overdensity. Consider a spherical object of constant dark matter density $\rho_b$ embedded in a much larger medium of density $\rho_c$ so that
\begin{equation}
\rho(r)=\left\{
  \begin{array}{l l}
   \rho_b,& r<R\\
    \rho_c,& r>R\\
  \end{array}\right. .
\end{equation}
When the object is static, equation (\ref{eq:eom}) becomes $\nabla^2\varphi=V_{\rm eff}(\varphi)_{\,,\varphi}$ and when the object is unscreened the field only differs from the exterior value by a small perturbation $\delta\varphi$ and so we can expand this to first order to find
\begin{equation}
 \nabla^2\delta\varphi\approx m_c^2\delta_\varphi+\frac{\bv(\varphi_0)\delta\rho}{\mpl},
\end{equation}
where $\delta\rho=\rho_b-\rho_c$ and $m_c^2$ is the mass of the field in the outer medium. Models with screening mechanisms typically have the property that the Compton wavelength of the field $m_c^{-1}$ is much greater than the size of the object and so one can neglect the mass term for the perturbation. In these models however, we have $m_c\ge m_{3/2}\ge1$ eV$\sim10^{28}$Mpc$^{-1}$ and therefore the Compton wavelength is incredibly small compared with typical galactic scales ($\sim\mathcal{O}(\textrm{Mpc})$). Therefore, the unscreened solution does not exist and all objects in the universe are screened.

\subsubsection{Supersymmetry Breaking}

Before looking at some specific examples of supersymmetric models we pause to discuss the effect of the dark-matter coupling on the supersymmetric properties of the model. Minimising (\ref{eq:veffPhi}) with respect to $\Phi$, one has
\begin{equation}\label{eq:qvoid}
 \left(\frac{K_{\Phi\Phi^\dagger\Phi^\dagger}}{K^2_{\Phi\Phi^\dagger}}-\frac{1}{K_{\Phi\Phi^\dagger}}\frac{\dd^2\hat{W}}{\dd\Phi^2}\right)\frac{\dd \hat{W}}{\dd \Phi}=\rc\frac{\dd A(\Phi)}{\dd \Phi}.
\end{equation}
The VEV of the dark matter scalars is $\langle\phi_\pm\rangle=0$ and so $F_\Phi=-\dd W/\dd \Phi=-\dd \hat{W}/\dd\Phi$. Any coupling to dark matter necessarily breaks supersymmetry at finite density. This is one of the new features of supersymmetric screened modified gravity; by secluding the dark sector from the observable one (up to supergravity breaking effects described above) the scale of supersymmetry breaking is not set by particle physics effects but rather by the ambient density and so our model is not plagued with issues such as the cosmological constant being associated with TeV scale breaking effects or detailed fine-tunings. That being said, this is far from a solution to the cosmological constant problem since we do not attempt to explain why the vacuum energy in the observable sector associated with QCD and electroweak symmetry breaking does not contribute to the cosmological dynamics. We also offer no explanation to the cancelling of the cosmological constant in the hidden sector.

\section{A Supersymmetric Chameleon}\label{sec:modelb}

Having found that symmetrons cannot operate within a supersymmetric framework we shall here construct a class of supersymmetric chameleon models with interesting, locally run-away potentials. Such run-away potentials are generally required to realise the screening mechanism (although this is not necessarily the case \cite{Brax:2010kv,Mota:2006fz}) and so it is instructive to pause to examine how one can construct these in supersymmetry. Of course there is no need of a chameleon screening mechanism here as the supergravity correction guarantees that the effects of the scalar field are essentially screened for all objects in the Universe. Here we will simply exemplify the construction of a chameleon model with a supersymmetric minimum. In this case, the cosmological evolution is directly influenced by the matter dependence of the minimum and the convergence of the field to the supersymmetric minimum when the matter density vanishes. We will see in the following section that these features allow one  to introduce a novel and supersymmetric way of generating a small cosmological constant whose presence can only be felt when the matter density falls below a certain threshold. As a result, this model encompasses all the characteristics of cosmological chameleons without the need to add a cosmological constant by hand in the scalar potential, something which would break supersymmetry explicitly. On the contrary, the cosmological constant and the resulting acceleration of the Universe at late time results entirely from a spontaneous breaking of global supersymmetry. 

\begin{figure}[ht]\centering
\includegraphics[width=0.5\textwidth]{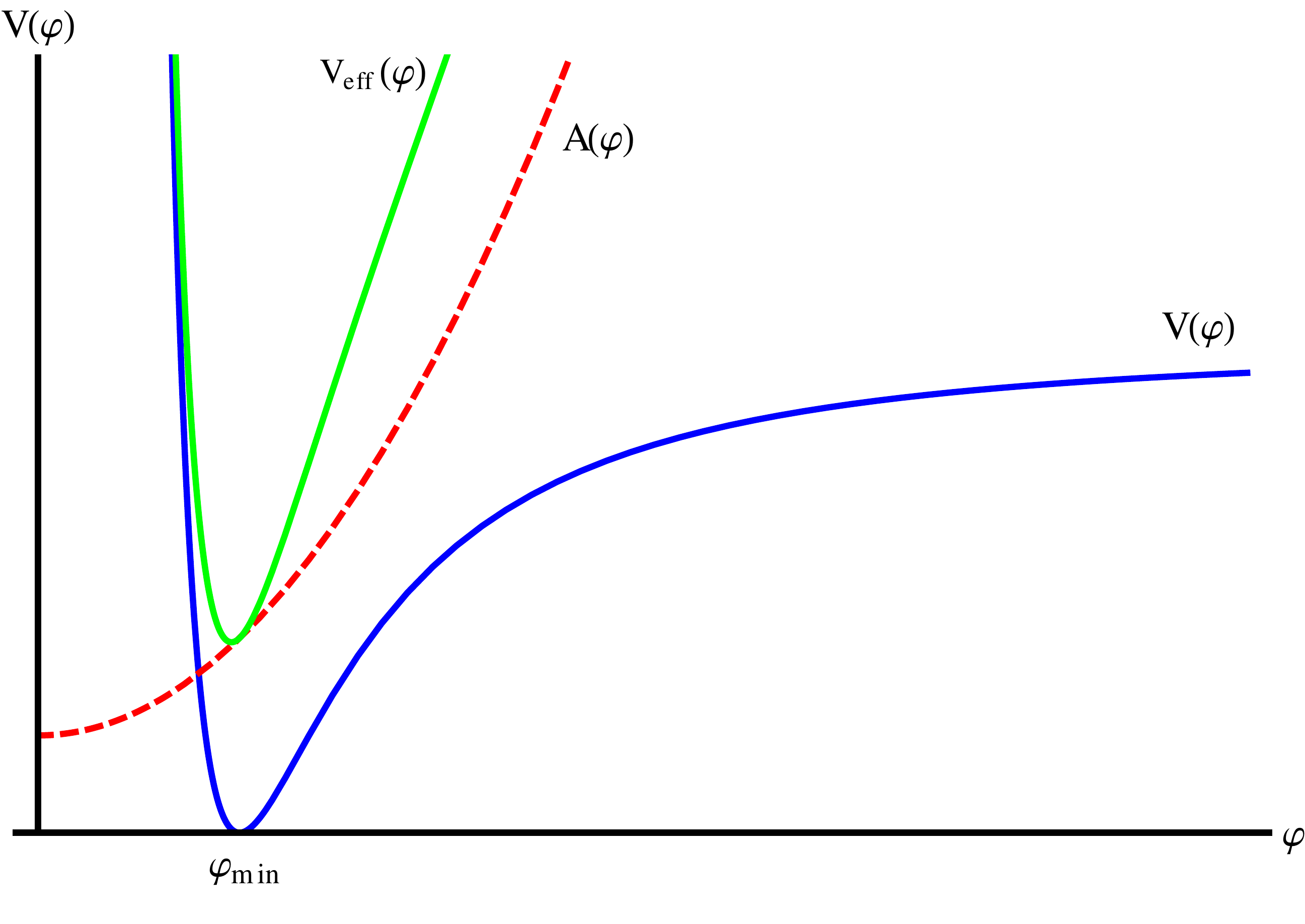}
\caption{The effective potential.}\label{fig:veff}
\end{figure}

First note that the scalar potential is given by (\ref{eq:VFpot}) where $\hat{W}$ typically varies as $\Phi^\alpha$ with $\alpha>0$. If one then wishes to have a run-away potential one needs to choose a K\"{a}hler potential of the form $(\Phi^\dagger\Phi)^\beta$ where $\beta>\alpha$. This immediately precludes the use of the canonical K\"{a}hler potential $K=\Phi^\dagger\Phi$, which in turn ensures that the field is not canonically normalised.
As a simple example, we will consider a particular form of the K\"{a}hler potential and superpotential which gives rise to a locally run-away potential at small field values and a supersymmetric minimum at larger ones. Hence we will find chameleon models where the true vacuum of the model, when the dark matter density vanishes, has a vanishing cosmological constant. To accommodate the late time acceleration of the Universe, we will introduce a new hybrid mechanism in the following section. Before doing this, constructing the model  can be achieved by the choice
\begin{align}
 \hat{W}&=\frac{\beta}{\sqrt{2}\alpha}\left(\frac{\Phi^\alpha}{\Lambda_0^{\alpha-3}}\right)+\frac{1}{\sqrt{2}}\left(\frac{\Phi^\beta}{\Lambda_2^{\beta-3}}\right),\\ \hat{K}(\Phi\Phi^\dagger)&=\frac{\Lambda_1^2}{2}\left(\frac{\Phi\Phi^\dagger}{\Lambda_1^2}\right)^\beta,\quad\textrm{and}\quad A(\Phi)=1+\frac{\Phi^\delta}{\Lambda_3^{\delta-1}}.
\end{align}
Minimising with respect to the angular field and defining the new mass scales
\begin{equation}\label{eq:scales}
 \Lambda^4\equiv\left(\frac{\Lambda_1}{\Lambda_2}\right)^{2\beta-2}{\Lambda_2}^4; \quad M^{n+4}=\left(\frac{\Lambda_1}{\Lambda_0}\right)^{2\beta-2}{\Lambda_0}^{n+4}
\end{equation}
with $n=2(\alpha-\beta)$, the F-term potential can be written
\begin{equation}\label{eq:Fpot}
 V_{\rm F}(\phi)=\left(\Lambda^2-\frac{M^{2+\frac{n}{2}}}{\phi^{\frac{n}{2}}}\right)^2=\Lambda^4\left[1-\left(\frac{\phi_{\rm min}}{\phi}\right)^\frac{n}{2}\right]^2,
\end{equation}
where
\begin{equation}\pmi=\left(\frac{M}{\Lambda}\right)^{\frac{4}{n}}M.
\end{equation}
This F-term potential has a supersymmetric minimum at $\phi=\pmi$ where $V(\pmi)=0$ and $\dd W/\dd \Phi=0$, however when $\phi\ll\pmi$ it reduces to the Ratra-Peebles potential
\begin{equation}
V_{\rm F}(\phi)\approx \Lambda^4 \left(\frac{\pmi}{\phi}\right)^n
\end{equation}
corresponding to a well-known dark energy model \cite{PhysRevD.37.3406}. When $\theta$ is set to its minimum, the  coupling function is
\begin{equation}\label{eq:AmodB}
 A(\phi)=1+\frac{g}{m{\Lambda_3}^{\delta-1}}\phi^{\delta}=1+\left(\frac{\phi}{\mu}\right)^{\delta};\quad \mu^\delta\equiv \frac{m\Lambda_3^{\delta-1}}{g},
\end{equation}
which can be cast into a more convenient form
\begin{equation}\label{eq:x}
A(\varphi)=1+x\left(\frac{\varphi}{\varphi_{\rm min}}\right)^{\frac{\delta}{\beta}};\quad x\equiv\frac{g\phi_{\rm min}^{\delta}}{m{\Lambda_3}^{\delta-1}}.
\end{equation}
The field $\phi$ is not canonically normalised, however the simple choice for the K\"{a}hler potential allows the normalised field to be found. It is given by
\begin{equation}
\varphi=\Lambda_1 \left(\frac{\phi}{\Lambda_1}\right)^\beta
\end{equation}
in which case the effective potential is
\begin{equation}\label{eq:bveff}
V_{\rm eff}(\varphi)=\Lambda^4\left[1-\left(\frac{\varphi_{\rm min}}{\varphi}\right)^{\frac{n}{2\beta}}\right]^2+\frac{m_{3/2}^2}{2} \varphi^2+ x\rc\left(\frac{\varphi}{\varphi_{\rm min}}\right)^{\frac{\delta}{\beta}}.
\end{equation}
This is shown schematically in figure (\ref{fig:veff})
where we have assumed that the supergravity correction can be neglected, i.e. $m_{3/2}^2 \phi_{\rm min}^2 \ll x \rho_\infty$.
Minimising the resulting potential one finds
\begin{equation}\label{eq:phimineq}
 \left(\frac{\varphi_{\rm min}}{\varphi}\right)^{\frac{n+\delta}{\beta}}-\left(\frac{\varphi_{\rm min}}{\varphi}\right)^{\frac{n+2\delta}{2\beta}}=\frac{\rc}{\rho_\infty},
\end{equation}
where
\begin{equation}\label{eq:rhoinf}
 \rho_\infty\equiv\rc^0(1+z_\infty)^3\equiv\frac{n\Lambda^4}{\delta x}.
\end{equation}
Clearly when $\rc\gg\rho_\infty$ we have $\varphi\ll\varphi_{\rm min}$ whilst when $\rho\lsim\rho_\infty$ the field lies very close to its supersymmetric minimum. $\rho_\infty$ therefore sets the density above which supersymmetry is broken.

This model gives rise to some interesting cosmological dynamics. The case $\delta=1$ has been studied previously in the context of the supersymmetron\cite{Brax:2011qs,Brax:2011bh}\footnote{Following this discussion the reader should be aware that the supersymmetron is a chameleon and not a symmetron.} and it was found that a cosmological constant is needed to account for both the energy density in dark energy and have $w$, the equation of state parameter, close to $-1$. This is in fact a very special case since the scale $\Lambda_3$ is absent. When $\delta\ne
1$ the dynamics are far more interesting and we examine them in detail in \cite{future_susy}. Here we shall only note that it is possible to account for both the energy density in dark energy and have $w\approx-1$, however this scenario predicts large deviations from the GR prediction for the matter power spectrum and so a cosmological constant is again required.

\section{Supersymmetric Hybrid Dark Energy}\label{sec:cc}

The model presented in the previous section requires a cosmological constant to match its predictions to current observations. Furthermore, it has been argued recently \cite{Wang:2012kj} that screened modified gravity cannot account for the acceleration of the universe without a cosmological constant. One would therefore expect supersymmetric theories to have the same requirement. Unlike more phenomenological theories however, the inclusion of such a cosmological constant at the level of the action is not so trivial. Global supersymmetry is necessarily broken if the vacuum energy is positive and so any cosmological constant must arise through the dynamics and will necessarily break supersymmetry. One could rely on supergravity breaking to generate such a term, however its magnitude is of order $m_{3/2}^2\mpl^2$, which is much greater than the observed value of $10^{-12}$ eV$^{4}$ and such contributions generally need to be fine-tuned away. In this section we will present a method by which a more natural low-energy scale cosmological constant can be generated via the cosmological dynamics of the field by coupling it to two chiral scalar fields charged under a $\mathrm{U}(1)$ gauge group using a hybrid-type mechanism.

We begin by adding a new coupling of the chameleon $\Phi$ to two $\mathrm{U}(1)$ charged chiral superfields $\Pi_\pm=\pi_\pm+\ldots$ and a $\mathrm{U}(1)$ gauge vector multiplet $X$ with Fayet-Illiopoulos term $\xi^2$. This is done by adding the term
\begin{equation}
 W_\pi=g^\prime\Phi\Pi_+\Pi_-
\end{equation}
to the superpotential (\ref{eq:suppotgen}). The introduction of this coupling greatly complicates the scalar potential, however when $\langle\pi_-\rangle=0$ our original effective potential (\ref{eq:bveff}) is recovered and a new potential for $\pi_+\equiv|\pi_+|$ coming from the D-term, the Fayet-Illiopoulos term and this coupling:
\begin{equation}\label{eq:D-term}
 V(\pi_+)=\frac{1}{2}\left(q\pi_+^2-\xi^2\right)^2+{g^{\prime}}^2\phi^2\pi_+^2.
\end{equation}
It can be shown \cite{future_susy} that $\langle\pi_-\rangle=0$ is indeed a stable minimum for the chameleon model of section (\ref{sec:modelb}). The effective mass of the charged field is $m_{\pi_+}^2=\gp^2\phi^2-q^2\xi^2$. In theories of screened modified gravity, including the model presented above \cite{future_susy}, the minimum of the effective potential is an attractor and the field tracks its position throughout its cosmological evolution \cite{Brax:2004qh,Brax:2012gr}. The mass of $\pi_+$ is then negative at early times when the field value is small but increases as the field evolves towards its supersymmetric minimum. We would therefore expect the shape of the potential (\ref{eq:D-term}) to change with the cosmological evolution of $\phi$. Indeed, if one minimises (\ref{eq:D-term}) one finds $\pi_+=0$ when $m_{\pi_+}\ge0$, corresponding to the restoration of the $\mathrm{U}(1)$ symmetry. When this is the case, one has $V(\pi_+=0)=\xi^4/2$ and so at late times the Fayet-Illiopoulos term plays the role of a cosmological constant. At earlier times, minimising the potential with respect to $\pi_+$ results in corrections to the effective potential for $\phi$. Exactly how (or indeed if at all) these corrections effect the model dynamics is highly model-dependent and requires a detailed numerical analysis. We will postpone any discussion of this to a follow-up paper \cite{future_susy}. Here we will only remark that in general it is possible to find a sensible (in the sense that no numbers are fine-tuned to ridiculous values) region of parameter space where the dynamics of the field are not disrupted by these corrections and the model predictions can be matched with current observations.

On this note, in order to match the presently observed density in dark energy we must set $\xi\sim10^{-3}$ eV. This may appear just as fine-tuned as any other quintessence model, however Fayet-Illiopoulos terms are far more stable under quantum corrections. When supersymmetry is unbroken there is no renormalisation and when it is broken there is only a logarithmic running \cite{Jack:1999zs} and so the cosmological constant is essentially uncorrected. Whereas there is no natural mechanism to set the specific value of $\xi$ within our framework, if one can find a more UV complete theory where such a scale naturally emerges then its value is preserved at low energies. One must also remember that this small value of the cosmological constant makes sense only if any other contribution coming from the matter and supersymmetry breaking sectors vanish. We do not attempt to address this issue here.

\section{Conclusions}\label{sec:concs}

We have presented a framework for embedding scalar-tensor theories into global $\mathcal{N}=1$ supersymmetry. These theories are IR modifications of gravity and so such a bottom-up approach is sensible. We have examined the new features this embedding has introduced and in particular, with the exception of a very small class of K\"{a}hler potentials, have found that any theory with underlying supergravity breaking, regardless of the origin of the effective potential, is so efficiently screened that there are no unscreened objects in the universe. Therefore we conclude that any scalar-tensor screening theory with origins in supergravity has no observational signatures due to fifth-forces. Only no-scale K\"{a}hler potentials with the isometry group (\ref{eq:manifod}) can evade this result and it is these model which one must investigate if one wishes to search for a non-trivial UV completion. Additionally, we have found that a contribution to the scalar mass of order $m_{3/2}$ is enough to obliterate the canonical and generalised symmetron mechanisms.

By secluding the dark sector from the observable sector we have circumvented TeV scale supersymmetry breaking effects on dark matter and dark energy. Indeed, we have shown that any supersymmetric model of screened modified gravity necessarily breaks supersymmetry at finite densities. The scale of this breaking is set by the parameters in the dark sector alone and can therefore be eV scale or below.

We have discussed the construction of supersymmetric chameleons and have argued that run-away type potentials can only be realised if one takes the K\"{a}hler potential to be of a higher degree in $\Phi^\dagger\Phi$ than the superpotential is in $\Phi$. We have illustrated this by constructing a specific class of models which are locally run-away but have a supersymmetric minimum at some large field value. At zero density supersymmetry is restored and at high densities it is broken by the matter coupling, however this class of models has the interesting feature that at small densities (exactly how small is parameter dependent) supersymmetry is approximately unbroken. In a follow-up work \cite{future_susy} we will investigate the dynamics of these models in detail.

Chameleon models generically require a cosmological constant, which cannot be present in supersymmetry at the level of the action. In order to address this issue we have introduced a novel mechanism where a coupling of the field to two $\mathrm{U}(1)$ charged scalars drives their mass from initially negative to positive values as the cosmological field tracks its (increasing in field space) density-dependent minimum. At some threshold density the mass is zero and the $ \mathrm{U}(1)$ symmetry is restored, driving the VEV of the charged scalars to zero and leaving only a Fayet-Illiopoulos term, which acts as a cosmological constant.

The exact value of the cosmological constant is not set by our low-energy description owing to the arbitrariness of the FI term and we must fix it to $10^{-3}$ eV in order to match the present-day dark energy density. FI terms do not receive large quantum corrections and run logarithmically at most. Our model is therefore more attractive than quintessence where the scalar VEV receives corrections from decoupling particle species. Furthermore, if one could find some natural mechanism for determining the value of the FI term from a more UV complete theory, for example, if it emerges as the ratio of  one large mass scale to another, even larger larger mass scale, then this value would be preserved through energy scales down to the dark energy one.

It seems that a supersymmetric extension of screened modified gravity theories brings with it new features and challenges to overcome. In general, any model which can screen does so to such an extent that there are no unscreened objects in the universe. This high degree of screening means that one would not expect any effects on non-linear structure formation such as spherical collapse or cluster abundances. Additionally, the extremely short range of the force also precludes any local observations resulting from fifth-forces originating from the coupling to matter\footnote{It is well known that a coupling between the field and photons is generated when there are heavy fermions present in the theory \cite{Brax:2010uq}, however we do not investigate this in this work. This investigation is left for a future analysis.}. The need for a cosmological constant makes the construction of a cosmologically viable model difficult within the framework of supersymmetry and one must find some mechanism by which it appears dynamically. Despite all this, we have successfully embedded these screening mechanisms into a supersymmetric framework and have introduced a novel, hybrid-type mechanism by which a stable (in that it is not overly sensitive to quantum corrections) cosmological constant can be generated dynamically in a manner consistent with the underlying supersymmetry.

Finally, one may wonder how the strong constraints on the supersymmetric embeddings of screened models could be relaxed and a model with more observation signatures could be constructed. This would certainly require the existence of no-scale models with the isometry group (\ref{eq:manifod}) and the presence of SUSY-flat directions so as to address the quantum corrections more formally. The construction of such a scenario is a challenging prospect and the framework we have presented here is a powerful tool with which to probe potential UV completions. This is only the first stage of a long process towards unifying these models with more fundamental theories.

\section*{Acknowledgements}

We are grateful to Neil Barnaby, Daniel Baumann and Eugene Lim for helpful discussions. JS and ACD (in part) is supported by the STFC.

\bibliographystyle{model1-num-names}
\bibliography{ref}

\begin{thebibliography}{40}
\expandafter\ifx\csname natexlab\endcsname\relax\def\natexlab#1{#1}\fi
\providecommand{\bibinfo}[2]{#2}
\ifx\xfnm\relax \def\xfnm[#1]{\unskip,\space#1}\fi
\bibitem[{Perlmutter et~al.(1999)}]{Perlmutter:1998np}
\bibinfo{author}{S.~Perlmutter}, et~al.,
\newblock \bibinfo{title}{{Measurements of Omega and Lambda from 42 high
  redshift supernovae}},
\newblock \bibinfo{journal}{Astrophys.J.} \bibinfo{volume}{517}
  (\bibinfo{year}{1999}) \bibinfo{pages}{565--586}.
\bibitem[{Riess et~al.(1998)}]{Riess:1998cb}
\bibinfo{author}{A.~G. Riess}, et~al.,
\newblock \bibinfo{title}{{Observational evidence from supernovae for an
  accelerating universe and a cosmological constant}},
\newblock \bibinfo{journal}{Astron.J.} \bibinfo{volume}{116}
  (\bibinfo{year}{1998}) \bibinfo{pages}{1009--1038}.
\bibitem[{Copeland et~al.(2006)Copeland, Sami, and Tsujikawa}]{Copeland:2006wr}
\bibinfo{author}{E.~J. Copeland}, \bibinfo{author}{M.~Sami},
  \bibinfo{author}{S.~Tsujikawa},
\newblock \bibinfo{title}{{Dynamics of dark energy}},
\newblock \bibinfo{journal}{Int.J.Mod.Phys.} \bibinfo{volume}{D15}
  (\bibinfo{year}{2006}) \bibinfo{pages}{1753--1936}.
\bibitem[{Weinberg(1965)}]{Weinberg:1965rz}
\bibinfo{author}{S.~Weinberg},
\newblock \bibinfo{title}{{Photons and gravitons in perturbation theory:
  Derivation of Maxwell's and Einstein's equations}},
\newblock \bibinfo{journal}{Phys.Rev.} \bibinfo{volume}{138}
  (\bibinfo{year}{1965}) \bibinfo{pages}{B988--B1002}.
\bibitem[{Jain and Khoury(2010)}]{Jain:2010ka}
\bibinfo{author}{B.~Jain}, \bibinfo{author}{J.~Khoury},
\newblock \bibinfo{title}{{Cosmological Tests of Gravity}},
\newblock \bibinfo{journal}{Annals Phys.} \bibinfo{volume}{325}
  (\bibinfo{year}{2010}) \bibinfo{pages}{1479--1516}.
\bibitem[{Khoury and Weltman(2004{\natexlab{a}})}]{Khoury:2003aq}
\bibinfo{author}{J.~Khoury}, \bibinfo{author}{A.~Weltman},
\newblock \bibinfo{title}{{Chameleon fields: Awaiting surprises for tests of
  gravity in space}},
\newblock \bibinfo{journal}{Phys.Rev.Lett.} \bibinfo{volume}{93}
  (\bibinfo{year}{2004}{\natexlab{a}}) \bibinfo{pages}{171104}.
\bibitem[{Khoury and Weltman(2004{\natexlab{b}})}]{Khoury:2003rn}
\bibinfo{author}{J.~Khoury}, \bibinfo{author}{A.~Weltman},
\newblock \bibinfo{title}{{Chameleon Cosmology}},
\newblock \bibinfo{journal}{Phys. Rev.} \bibinfo{volume}{D69}
  (\bibinfo{year}{2004}{\natexlab{b}}) \bibinfo{pages}{044026}.
\bibitem[{Brax et~al.(2004)Brax, van~de Bruck, Davis, Khoury, and
  Weltman}]{Brax:2004qh}
\bibinfo{author}{P.~Brax}, \bibinfo{author}{C.~van~de Bruck},
  \bibinfo{author}{A.-C. Davis}, \bibinfo{author}{J.~Khoury},
  \bibinfo{author}{A.~Weltman},
\newblock \bibinfo{title}{{Detecting dark energy in orbit: The cosmological
  chameleon}},
\newblock \bibinfo{journal}{Phys. Rev.} \bibinfo{volume}{D70}
  (\bibinfo{year}{2004}) \bibinfo{pages}{123518}.
\bibitem[{Hinterbichler and Khoury(2010)}]{Hinterbichler:2010es}
\bibinfo{author}{K.~Hinterbichler}, \bibinfo{author}{J.~Khoury},
\newblock \bibinfo{title}{{Symmetron Fields: Screening Long-Range Forces
  Through Local Symmetry Restoration}},
\newblock \bibinfo{journal}{Phys. Rev. Lett.} \bibinfo{volume}{104}
  (\bibinfo{year}{2010}) \bibinfo{pages}{231301}.
\bibitem[{Brax et~al.(2010)Brax, van~de Bruck, Davis, and Shaw}]{Brax:2010gi}
\bibinfo{author}{P.~Brax}, \bibinfo{author}{C.~van~de Bruck},
  \bibinfo{author}{A.-C. Davis}, \bibinfo{author}{D.~Shaw},
\newblock \bibinfo{title}{{The Dilaton and Modified Gravity}},
\newblock \bibinfo{journal}{Phys. Rev.} \bibinfo{volume}{D82}
  (\bibinfo{year}{2010}) \bibinfo{pages}{063519}.
\bibitem[{Wang et~al.(2012)Wang, Hui, and Khoury}]{Wang:2012kj}
\bibinfo{author}{J.~Wang}, \bibinfo{author}{L.~Hui},
  \bibinfo{author}{J.~Khoury},
\newblock \bibinfo{title}{{No-Go Theorems for Generalized Chameleon Field
  Theories}}  (\bibinfo{year}{2012}).
\bibitem[{Upadhye et~al.(2012)Upadhye, Hu, and Khoury}]{Upadhye:2012vh}
\bibinfo{author}{A.~Upadhye}, \bibinfo{author}{W.~Hu},
  \bibinfo{author}{J.~Khoury},
\newblock \bibinfo{title}{{Quantum Stability of Chameleon Field Theories}}
  (\bibinfo{year}{2012}).
\bibitem[{Gannouji et~al.(2012)Gannouji, Sami, and Thongkool}]{Gannouji:2012iy}
\bibinfo{author}{R.~Gannouji}, \bibinfo{author}{M.~Sami},
  \bibinfo{author}{I.~Thongkool},
\newblock \bibinfo{title}{{Generic f(R) theories and classicality of their
  scalarons}}  (\bibinfo{year}{2012}).
\bibitem[{Brax and Martin(2006)}]{Brax:2006kg}
\bibinfo{author}{P.~Brax}, \bibinfo{author}{J.~Martin},
\newblock \bibinfo{title}{{The SUGRA Quintessence Model Coupled to the MSSM}},
\newblock \bibinfo{journal}{JCAP} \bibinfo{volume}{0611} (\bibinfo{year}{2006})
  \bibinfo{pages}{008}.
\bibitem[{Brax and Martin(2007{\natexlab{a}})}]{Brax:2006dc}
\bibinfo{author}{P.~Brax}, \bibinfo{author}{J.~Martin},
\newblock \bibinfo{title}{{Dark Energy and the MSSM}},
\newblock \bibinfo{journal}{Phys. Rev.} \bibinfo{volume}{D75}
  (\bibinfo{year}{2007}{\natexlab{a}}) \bibinfo{pages}{083507}.
\bibitem[{Brax and Martin(2007{\natexlab{b}})}]{Brax:2006np}
\bibinfo{author}{P.~Brax}, \bibinfo{author}{J.~Martin},
\newblock \bibinfo{title}{{Moduli fields as quintessence and the chameleon}},
\newblock \bibinfo{journal}{Phys. Lett.} \bibinfo{volume}{B647}
  (\bibinfo{year}{2007}{\natexlab{b}}) \bibinfo{pages}{320--329}.
\bibitem[{Hinterbichler et~al.(2011)Hinterbichler, Khoury, and
  Nastase}]{Hinterbichler:2010wu}
\bibinfo{author}{K.~Hinterbichler}, \bibinfo{author}{J.~Khoury},
  \bibinfo{author}{H.~Nastase},
\newblock \bibinfo{title}{{Towards a UV Completion for Chameleon Scalar
  Theories}},
\newblock \bibinfo{journal}{JHEP} \bibinfo{volume}{1103} (\bibinfo{year}{2011})
  \bibinfo{pages}{061}.
\bibitem[{Conlon and Pedro(2011)}]{Conlon:2010jq}
\bibinfo{author}{J.~P. Conlon}, \bibinfo{author}{F.~G. Pedro},
\newblock \bibinfo{title}{{Moduli-Induced Vacuum Destabilisation}},
\newblock \bibinfo{journal}{JHEP} \bibinfo{volume}{1105} (\bibinfo{year}{2011})
  \bibinfo{pages}{079}.
\bibitem[{Brax and Davis(2012)}]{Brax:2011qs}
\bibinfo{author}{P.~Brax}, \bibinfo{author}{A.-C. Davis},
\newblock \bibinfo{title}{{Supersymmetron}},
\newblock \bibinfo{journal}{Phys.Lett.} \bibinfo{volume}{B707}
  (\bibinfo{year}{2012}) \bibinfo{pages}{1--7}.
\bibitem[{Brax et~al.(????)Brax, Davis, and Sakstein}]{future_susy}
\bibinfo{author}{P.~Brax}, \bibinfo{author}{A.-C. Davis},
  \bibinfo{author}{J.~Sakstein},
\newblock \bibinfo{title}{{Dynamics of Supersymmetric Chameleons}},
\newblock \bibinfo{journal}{In Preperation}  (????).
\bibitem[{Brax et~al.(2010)Brax, van~de Bruck, Mota, Nunes, and
  Winther}]{Brax:2010kv}
\bibinfo{author}{P.~Brax}, \bibinfo{author}{C.~van~de Bruck},
  \bibinfo{author}{D.~F. Mota}, \bibinfo{author}{N.~J. Nunes},
  \bibinfo{author}{H.~A. Winther},
\newblock \bibinfo{title}{{Chameleons with Field Dependent Couplings}},
\newblock \bibinfo{journal}{Phys. Rev.} \bibinfo{volume}{D82}
  (\bibinfo{year}{2010}) \bibinfo{pages}{083503}.
\bibitem[{Brax et~al.(2008)Brax, van~de Bruck, Davis, and Shaw}]{Brax:2008hh}
\bibinfo{author}{P.~Brax}, \bibinfo{author}{C.~van~de Bruck},
  \bibinfo{author}{A.-C. Davis}, \bibinfo{author}{D.~J. Shaw},
\newblock \bibinfo{title}{{f(R) Gravity and Chameleon Theories}},
\newblock \bibinfo{journal}{Phys.Rev.} \bibinfo{volume}{D78}
  (\bibinfo{year}{2008}) \bibinfo{pages}{104021}.
\bibitem[{Brax et~al.(2012)Brax, Davis, Li, and Winther}]{Brax:2012gr}
\bibinfo{author}{P.~Brax}, \bibinfo{author}{A.-C. Davis},
  \bibinfo{author}{B.~Li}, \bibinfo{author}{H.~A. Winther},
\newblock \bibinfo{title}{{A Unified Description of Screened Modified Gravity}}
   (\bibinfo{year}{2012}).
\bibitem[{Khoury(2010)}]{Khoury:2010xi}
\bibinfo{author}{J.~Khoury},
\newblock \bibinfo{title}{{Theories of Dark Energy with Screening Mechanisms}}
  (\bibinfo{year}{2010}).
\bibitem[{Nilles(1984)}]{Nilles:1983ge}
\bibinfo{author}{H.~P. Nilles},
\newblock \bibinfo{title}{{Supersymmetry, Supergravity and Particle Physics}},
\newblock \bibinfo{journal}{Phys.Rept.} \bibinfo{volume}{110}
  (\bibinfo{year}{1984}) \bibinfo{pages}{1--162}.
\bibitem[{Hinterbichler et~al.(2011)Hinterbichler, Khoury, Levy, and
  Matas}]{Hinterbichler:2011ca}
\bibinfo{author}{K.~Hinterbichler}, \bibinfo{author}{J.~Khoury},
  \bibinfo{author}{A.~Levy}, \bibinfo{author}{A.~Matas},
\newblock \bibinfo{title}{{Symmetron Cosmology}}  (\bibinfo{year}{2011}).
\bibitem[{Brignole et~al.(1994)Brignole, Ibanez, and Munoz}]{Brignole:1993dj}
\bibinfo{author}{A.~Brignole}, \bibinfo{author}{L.~E. Ibanez},
  \bibinfo{author}{C.~Munoz},
\newblock \bibinfo{title}{{Towards a theory of soft terms for the
  supersymmetric Standard Model}},
\newblock \bibinfo{journal}{Nucl.Phys.} \bibinfo{volume}{B422}
  (\bibinfo{year}{1994}) \bibinfo{pages}{125--171}.
\bibitem[{Brignole et~al.(1997)Brignole, Ibanez, and Munoz}]{Brignole:1997dp}
\bibinfo{author}{A.~Brignole}, \bibinfo{author}{L.~E. Ibanez},
  \bibinfo{author}{C.~Munoz},
\newblock \bibinfo{title}{{Soft supersymmetry-breaking terms from supergravity
  and superstring models}}  (\bibinfo{year}{1997}).
\bibitem[{Farquet and Scrucca(2012)}]{Farquet:2012cs}
\bibinfo{author}{D.~Farquet}, \bibinfo{author}{C.~A. Scrucca},
\newblock \bibinfo{title}{{Scalar geometry and masses in Calabi-Yau string
  models}},
\newblock \bibinfo{journal}{JHEP} \bibinfo{volume}{1209} (\bibinfo{year}{2012})
  \bibinfo{pages}{025}.
\bibitem[{Kachru et~al.(2003)Kachru, Kallosh, Linde, and
  Trivedi}]{Kachru:2003aw}
\bibinfo{author}{S.~Kachru}, \bibinfo{author}{R.~Kallosh},
  \bibinfo{author}{A.~D. Linde}, \bibinfo{author}{S.~P. Trivedi},
\newblock \bibinfo{title}{{De Sitter vacua in string theory}},
\newblock \bibinfo{journal}{Phys.Rev.} \bibinfo{volume}{D68}
  (\bibinfo{year}{2003}) \bibinfo{pages}{046005}.
\bibitem[{Hui et~al.(2009)Hui, Nicolis, and Stubbs}]{Hui:2009kc}
\bibinfo{author}{L.~Hui}, \bibinfo{author}{A.~Nicolis},
  \bibinfo{author}{C.~Stubbs},
\newblock \bibinfo{title}{{Equivalence Principle Implications of Modified
  Gravity Models}},
\newblock \bibinfo{journal}{Phys.Rev.} \bibinfo{volume}{D80}
  (\bibinfo{year}{2009}) \bibinfo{pages}{104002}.
\bibitem[{Davis et~al.(2012)Davis, Lim, Sakstein, and
  Shaw}]{PhysRevD.85.123006}
\bibinfo{author}{A.-C. Davis}, \bibinfo{author}{E.~A. Lim},
  \bibinfo{author}{J.~Sakstein}, \bibinfo{author}{D.~J. Shaw},
\newblock \bibinfo{title}{Modified gravity makes galaxies brighter},
\newblock \bibinfo{journal}{Phys. Rev. D} \bibinfo{volume}{85}
  (\bibinfo{year}{2012}) \bibinfo{pages}{123006}.
\bibitem[{Chang and Hui(2010)}]{Chang:2010xh}
\bibinfo{author}{P.~Chang}, \bibinfo{author}{L.~Hui},
\newblock \bibinfo{title}{{Stellar Structure and Tests of Modified Gravity}}
  (\bibinfo{year}{2010}).
\bibitem[{Jain et~al.(2012)Jain, Vikram, and Sakstein}]{Jain:2012tn}
\bibinfo{author}{B.~Jain}, \bibinfo{author}{V.~Vikram},
  \bibinfo{author}{J.~Sakstein},
\newblock \bibinfo{title}{{Astrophysical Tests of Modified Gravity: Constraints
  from Distance Indicators in the Nearby Universe}}  (\bibinfo{year}{2012}).
\bibitem[{Jain and VanderPlas(2011)}]{Jain:2011ji}
\bibinfo{author}{B.~Jain}, \bibinfo{author}{J.~VanderPlas},
\newblock \bibinfo{title}{{Tests of Modified Gravity with Dwarf Galaxies}},
\newblock \bibinfo{journal}{JCAP} \bibinfo{volume}{1110} (\bibinfo{year}{2011})
  \bibinfo{pages}{032}.
\bibitem[{Mota and Shaw(2007)}]{Mota:2006fz}
\bibinfo{author}{D.~F. Mota}, \bibinfo{author}{D.~J. Shaw},
\newblock \bibinfo{title}{{Evading equivalence principle violations,
  astrophysical and cosmological constraints in scalar field theories with a
  strong coupling to matter}},
\newblock \bibinfo{journal}{Phys. Rev.} \bibinfo{volume}{D75}
  (\bibinfo{year}{2007}) \bibinfo{pages}{063501}.
\bibitem[{Ratra and Peebles(1988)}]{PhysRevD.37.3406}
\bibinfo{author}{B.~Ratra}, \bibinfo{author}{P.~J.~E. Peebles},
\newblock \bibinfo{title}{Cosmological consequences of a rolling homogeneous
  scalar field},
\newblock \bibinfo{journal}{Phys. Rev. D} \bibinfo{volume}{37}
  (\bibinfo{year}{1988}) \bibinfo{pages}{3406--3427}.
\bibitem[{Brax et~al.(2011)Brax, Davis, and Winther}]{Brax:2011bh}
\bibinfo{author}{P.~Brax}, \bibinfo{author}{A.-C. Davis},
  \bibinfo{author}{H.~A. Winther},
\newblock \bibinfo{title}{{The Cosmological Supersymmetron}}
  (\bibinfo{year}{2011}).
\bibitem[{Jack and Jones(2000)}]{Jack:1999zs}
\bibinfo{author}{I.~Jack}, \bibinfo{author}{D.~Jones},
\newblock \bibinfo{title}{{Renormalization of the Fayet-Iliopoulos D term}},
\newblock \bibinfo{journal}{Phys.Lett.} \bibinfo{volume}{B473}
  (\bibinfo{year}{2000}) \bibinfo{pages}{102--108}.
\bibitem[{Brax et~al.(2011)Brax, Burrage, Davis, Seery, and
  Weltman}]{Brax:2010uq}
\bibinfo{author}{P.~Brax}, \bibinfo{author}{C.~Burrage}, \bibinfo{author}{A.-C.
  Davis}, \bibinfo{author}{D.~Seery}, \bibinfo{author}{A.~Weltman},
\newblock \bibinfo{title}{{Anomalous coupling of scalars to gauge fields}},
\newblock \bibinfo{journal}{Phys. Lett.} \bibinfo{volume}{B699}
  (\bibinfo{year}{2011}) \bibinfo{pages}{5--9}.

\end{thebibliography}
\end{document}